\documentclass [aps,prl,twocolumn,showpacs,nofootinbib,superscriptaddress,preprintnumbers]{revtex4}
\usepackage{epsfig}
\usepackage{graphicx}
\usepackage{amsmath}
\usepackage{mathrsfs}
\usepackage{bm}

\begin{document}

\preprint{DESY~19--097\hspace{14.cm} ISSN 0418--9833$\hphantom{XXXXX}$}
\preprint{June 2019\hspace{18.4cm}}

\boldmath
\title{Deciphering the $X(3872)$ via its Polarization in Prompt Production at
the CERN LHC}
\unboldmath
\author{Mathias Butenschoen}
\affiliation{{II.} Institut f\"ur Theoretische Physik, Universit\"at Hamburg,
Luruper Chaussee 149, 22761 Hamburg, Germany}
\author{Zhi-Guo He}
\affiliation{{II.} Institut f\"ur Theoretische Physik, Universit\"at Hamburg,
Luruper Chaussee 149, 22761 Hamburg, Germany}
\author{Bernd A. Kniehl}
\affiliation{{II.} Institut f\"ur Theoretische Physik, Universit\"at Hamburg,
Luruper Chaussee 149, 22761 Hamburg, Germany}
\date{\today}
\begin{abstract}\vspace{5mm}
  Based on the hypothesis that the $X(3872)$ exotic hadron is a mixture of
  $\chi_{c1}(2P)$ and other states and that its prompt hadroproduction
  predominately proceeds via its $\chi_{c1}(2P)$ component, we calculate the
  prompt-$X(3872)$ polarization at the CERN LHC through next-to-leading order
  in $\alpha_s$ within the factorization formalism of nonrelativistic QCD,
  including both the color-singlet $^3\!P_1^{[1]}$ and color-octet
  $^3\!S_1^{[8]}$ $c\bar c$ Fock states.
  We also consider the polarization of the $J/\psi$ produced by the subsequent
  $X(3872)$ decay.
  We predict that, under ATLAS, CMS, and LHCb experimental conditions, the
  $X(3872)$ is largely longitudinally polarized, while the $J/\psi$ is largely
  transversely polarized.
  We propose that the LHC experiments perform such polarization measurements to
  pin down the nature of the $X(3872)$ and other $X$, $Y$, $Z$ exotic states
  with nonzero spin.
\end{abstract}


\maketitle
The discovery of the $X(3872)$ by Belle in 2003
\cite{Choi:2003ue}, which soon afterward was confirmed by CDF
\cite{Acosta:2003zx}, D0 \cite{Abazov:2004kp}, and BaBar \cite{Aubert:2004ns},
triggered the renaissance of hadron spectroscopy. 
Ever since then, many charmonium or charmonium-like states, named $X,Y,Z$, were
discovered, and numerous theoretical studies were devoted to reveal their
nature.
We refer to Refs.~\cite{Olsen:2017bmm,Esposito:2016noz} as the latest reviews
of experimental results and theoretical approaches, respectively.
Among the $X,Y,Z$ hadrons, the $X(3872)$ is the top highlighted state.
CDF \cite{Abulencia:2005zc} and LHCb \cite{Aaij:2013zoa}
have established the $J^{PC}=1^{++}$ quantum numbers of the $X(3872)$, and the
very precise world average of its mass is $m_X=3871.69\pm0.17$~MeV
\cite{Tanabashi:2018oca}.
On the theory side, however, we are still far away from a convincing, overall
picture to explain all the measurements.
The popular models on the market include charmonium \cite{Barnes:2003vb},
$D^{\ast0}\bar{D}^0/D^0\bar{D}^{\ast0}$ molecule \cite{Close:2003sg},
tetraquark \cite{Maiani:2004vq}, hybrid \cite{Li:2004sta}, or some
quantum-mechanical mixtures thereof; see, e.g., Ref.~\cite{Chen:2016qju} for
reviews.
At hadron colliders, the $X(3872)$ is most frequently produced promptly,
as has been observed by CDF \cite{Acosta:2003zx,Bauer:2004bc}
at the Fermilab Tevatron and by LHCb \cite{Aaij:2011sn}, CMS
\cite{Chatrchyan:2013cld}, and ATLAS \cite{Aaboud:2016vzw} at
the CERN LHC.
Besides mass spectrum and decay modes, prompt production provides a
complementary source of information on the nature of the $X(3872)$.
For example, in our previous work \cite{Butenschoen:2013pxa}, we showed that
the pure $\chi_{c1}(2P)$ option of the $X(3872)$ can be excluded by
analyzing its prompt hadroproduction rates in the framework of
nonrelativistic-QCD (NRQCD) factorization \cite{Bodwin:1994jh}, and we
predicted the $\chi_{c1}(2P)$ component to be around 30\% under the assumption
that the $X(3872)$ state is a quantum mechanical mixture of a $\chi_{c1}(2P)$
and a $D^{\ast0}\bar{D}^0/D^0\bar{D}^{\ast0}$ molecule as proposed in
Ref.~\cite{Meng:2005er}.
The study of $X(3872)$ prompt production within NRQCD factorization was
pioneered by Artoisenet and Braaten \cite{Artoisenet:2009wk}, who considered
the color octet (CO) contribution, due to the $c\bar{c}$ Fock state
$^3S_1^{[8]}$, at leading order (LO).
Prompt $X(3872)$ hadroproduction was also studied in the molecular picture, and
the cross section was found to greatly undershoot CDF data
\cite{Bignamini:2009sk}.
Although this problem could be remedied \cite{Albaladejo:2017blx} by properly
taking into account the rescattering mechanism \cite{Artoisenet:2009wk}, it
is still under debate if the molecular picture can adequately describe all the
experimental data \cite{Wang:2017gay}. 
Very recently, $X(3872)$ plus soft-pion production has been proposed to settle
this issue \cite{Braaten:2018eov}.

Despite a concerted experimental and theoretical endeavor during the past
decade, the quest for the ultimate classification of the $X(3872)$ and other
$X,Y,Z$ states remains one of the most tantalizing challenges of hadron
spectroscopy at the present time.
Since the total spin of the $X(3872)$ is 1, rather than 0, its
polarization in prompt production is expected to be rather sensitive to its
production mechanism and its internal structure.
Moreover, the $X(3872)$ has a considerable branching fraction to decay
into the $J/\psi$.
Under the assumption that the total spin of the charm quark pair is preserved
during the decay process, the polarization of the $J/\psi$ from the
decay of the prompt $X(3872)$ will help us to analyze the role of
the $c\bar{c}$ pair inside the $X(3872)$.
To decipher the as-yet inscrutable nature of the $X(3872)$, we thus
propose in this Letter to measure at the LHC its polarization and that of the
$J/\psi$ that springs from it.
Working in NRQCD factorization at next-to-leading order (NLO) in $\alpha_s$, we
provide here, for the first time, the respective theoretical predictions under
the 
likely assumption that the prompt hadroproduction of the $X(3872)$
proceeds predominately via 
the $\chi_{c1}(2P)$ component of its wave function at short distances.
By doing so, we correct a reproducible error, common to existing literature on
the NLO NRQCD treatment of $P$-wave heavy-quark pair polarization, related to the
implementation of phase space slicing \cite{Butenschoen:2012px,Gong:2012ug}.

The observation that NRQCD factorization at NLO fails to yield a coherent
description of the world data on $J/\psi$ yield and polarization
\cite{Butenschoen:2012px,Butenschoen:2014dra} may not affect our present
analysis.
In contrast to the $\chi_{cJ}$ case relevant here, the color singlet (CS)
contribution to direct $J/\psi$ hadroproduction has not yet unfolded its
leading power, proportional to $1/p_T^4$, at NLO \cite{Ma:2014svb}.
This will only happen at next-to-next-to-leading order, where a new
dominant CS production channel will open up, which will dynamically create an
enhancement that is likely to exceed the parametric $O(\alpha_s)$ by orders of
magnitude, with the potential to reconcile the $J/\psi$ world data with NRQCD
factorization.
Furthermore, the ${}^3P_J^{[8]}$ CO channel, which contributes to $S$, but not
$P$ wave quarkonium production, is potentially very sensitive to
next-to-next-to-leading-order
corrections, owing to a cancellation between LO and NLO corrections
\cite{Butenschoen:2012px,Butenschoen:2010rq}.
Finally, the $J/\psi$ polarization problem
\cite{Butenschoen:2012px,Butenschoen:2014dra} is reduced to a tolerable level
if one takes the point of view that data with $p_T<10~\mathrm{GeV}$ should be
disregarded to suppress contributions violating NRQCD factorization
\cite{Chao:2012iv}, the more so if the leading logarithms in $p_T^2/m_c^2$ are
resummed \cite{Bodwin:2014gia}.

Adopting the collinear parton model of QCD and NRQCD factorization, the spin
density matrix of the differential cross section of prompt $X(3872)$
hadroproduction can be evaluated as
\begin{eqnarray}\label{xs}
\lefteqn{\hspace{-0.4cm}d\sigma_{ij}(AB\to X(3872)+\mathrm{anything})=
\sum_{k,l,n} \int dxdy\,f_{k/A}(x)}
\nonumber\\
&&{}\hspace{-0.8cm}\times f_{l/B}(y)\,
d\hat{\sigma}_{ij}(k l\to c\overline{c}[n]+\mathrm{anything})\,
\langle\overline{\mathcal{O}}^{X(3872)}[n]\rangle,
\end{eqnarray}
where $f_{k/A}(x)$ is the density function (PDF) of parton $k$ with
longitudinal-momentum fraction $x$ inside hadron $A$,
$d\hat{\sigma}_{ij}(k l\to c\overline{c}[n]+\mathrm{anything})$ with
$i,j=0,\pm1$ is the spin density matrix element of the respective
partonic cross section, and
$\langle\overline{\mathcal{O}}^{X(3872)}[n]\rangle
=\langle\mathcal{O}^{\chi_{c1}(2P)}[n]\rangle
|\langle\chi_{c1}(2P)|X(3872)\rangle|^2$,
with $\langle\mathcal{O}^{\chi_{c1}(2P)}[n]\rangle$ being the long-distance
matrix element (LDME) of the $c\bar{c}$ Fock state $n$ inside the
$\chi_{c1}(2P)$ and $\langle\chi_{c1}(2P)|X(3872)\rangle$ being the overlap of
the physical $X(3872)$ and $\chi_{c1}(2P)$ wave functions.
At LO in $v^2$, where $v$ is the relative velocity in the
motion of the $c\bar{c}$ pair, we have $n={}^3P_1^{[1]},{}^3S_1^{[8]}$, where
${}^{2S+1}L_J$ refers to the spectroscopic notation and the label in brackets
indicates CS and CO configurations.
The evaluation of
$d\hat{\sigma}_{ij}(k l\to c\overline{c}[n]+\mathrm{anything})$ at NLO in NRQCD
proceeds as in Ref.~\cite{Butenschoen:2012px} upon the correction mentioned
above.
The production of polarized $J/\psi$'s via the feed down of promptly
hadroproduced $X(3872)$'s may be treated in one sweep, adopting the
formalism outlined in Ref.~\cite{He:2015gla}.
NLO NRQCD polarization studies for promptly hadroproduced $\chi_{c1}$'s and
$\chi_{c2}$'s may be found in Refs.~\cite{Bodwin:2014gia,Shao:2014fca}.

\begin{figure}
\begin{tabular}{cc}
\includegraphics[width=0.48\linewidth]{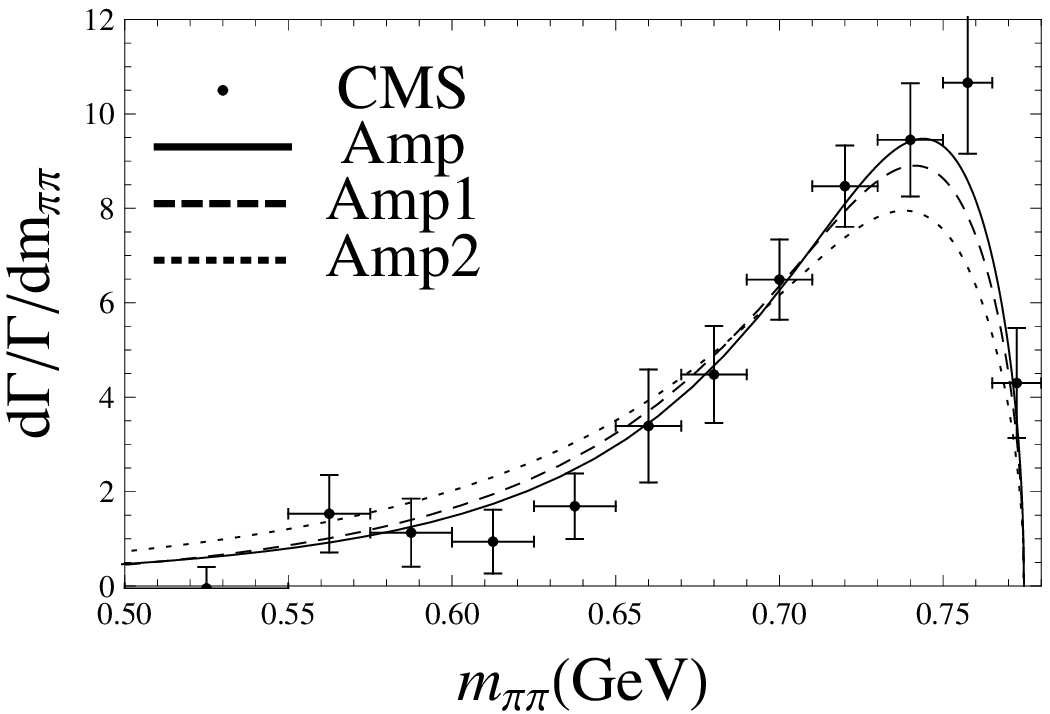}&
\includegraphics[width=0.48\linewidth]{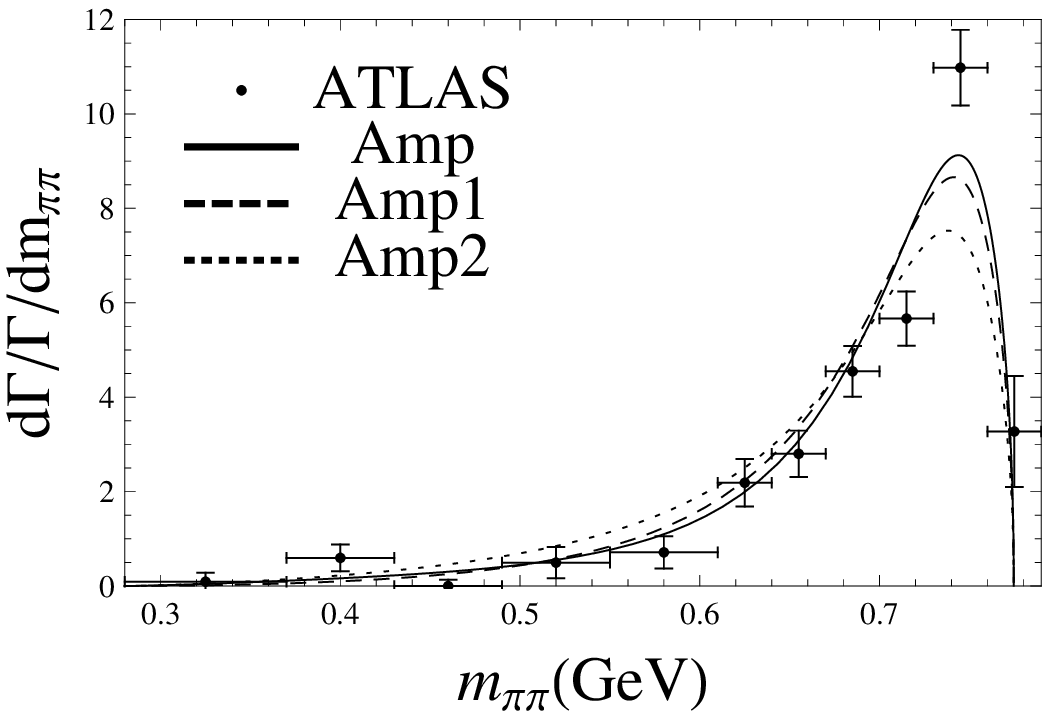}
\end{tabular}
\caption{The $m_{\pi\pi}$ distributions of
  $\Gamma(X(3872)\to J/\psi\pi^{+}\pi^{-})$ measured by CMS
  \cite{Chatrchyan:2013cld} (left panel) and ATLAS \cite{Aaboud:2016vzw} (right
  panel), normalized to unity across the experimental $m_{\pi\pi}$ ranges, are
  compared with our theoretical predictions based on $A^{1}_{\mu_1}$ (dashed
  lines), $A^{2}_{\mu_1}$ (dotted lines), and their linear combination in
  Eq.~(\ref{amprho}) with fitted value of $g$ (solid lines).}
\label{Fitpipi}
\end{figure}

The polarizations of the $X(3872)$ and $J/\psi$ can be measured by
analyzing the angular distributions of their decay products.
The $J/\psi$ is easily reconstructed by its decay to a lepton pair
$l^+l^-$, and the distribution in the polar angle $\theta$ of the
$l^{+}$ flight direction in the $J/\psi$ rest frame reads \cite{Lam:1978pu}
\begin{equation}\label{jsipol}
W_{\psi}(\theta)\propto 1+\lambda^{\psi}_{\theta}\cos^2\theta,
\end{equation}
where the polarization parameter
$\lambda^{\psi}_{\theta}=(\sigma_{11}^{\psi}-\sigma_{00}^{\psi})/
(\sigma_{11}^{\psi}+\sigma_{00}^{\psi})$
takes the values $0,\pm1$ if the $J/\psi$ is unpolarized and
totally transversely or longitudinally polarized, respectively.
The definition of $\theta$ depends on the choice of coordinate frame.
The helicity (HX) frame, in which the polar axis is chosen to point along the
$J/\psi$ flight direction in the center-of-mass (c.m.) frame of the collision,
and the Collins-Soper frame, in which the polar axis is defined as the
bisector of the two beam directions \cite{Collins:1977iv}, are most frequently
used experimentally.
For definiteness, we will use the HX frame throughout this Letter.
The counterpart of Eq.~(\ref{jsipol}) for the $X(3872)$ is not yet
available and will be derived in the following.
So far, all $X(3872)$ events of prompt hadroproduction have been reconstructed
through the $J/\psi\pi^{+}\pi^{-}$ decay channel.
CMS \cite{Chatrchyan:2013cld} and ATLAS \cite{Aaboud:2016vzw}
found that almost all the $\pi^{+}\pi^{-}$ pairs originate
from $\rho$ vector meson decay.
Because of this $\rho$ dominance, the partial decay amplitude of
$X(3872)\to J/\psi\pi^{+}\pi^{-}$ can be approximated by
\begin{eqnarray}
\lefteqn{\hspace{-0.5cm}\mathcal{M}(X(3872)\to J/\psi\pi^{+}\pi^{-})
=\mathcal{A}_{\mu}(X(3872)\to J/\psi \rho)}\nonumber\\
&&{}\times(-g^{\mu\nu}+p_\rho^{\mu}p_\rho^{\nu}/m^2_{\rho})\mathrm{BW}_\rho(p_\rho^2)
\mathcal{A}_{\nu}(\rho\to \pi^{+}\pi^{-}),
\label{ampX}
\end{eqnarray}
where $p_\rho$ is the four-momentum of the intermediate $\rho$, $m_{\rho}$ is
its mass, and  $\mathrm{BW}_{\rho}(p_\rho^2)$ is its propagator in
Breit-Wigner form.
As is well known \cite{Braaten:2005ai}, we have
$\mathcal{A}^{\mu}(\rho\to \pi^{+}\pi^{-})
=f_{\rho\pi\pi}(p_{\pi^+}^{\mu}-p_{\pi^-}^{\mu})$, where
$f_{\rho\pi\pi}$ is a hadronic coupling constant.
On the other hand, $CPT$ conservation and Lorentz covariance restrict
$\mathcal{A}_{\mu}(X(3872)\to J/\psi \rho)$ to be a linear combination of
$\mathcal{A}_{\mu}^{1}= \varepsilon_{\mu\alpha\beta\gamma} 
\epsilon_{X}^{\alpha}\epsilon_{\psi}^{\ast\beta}p_{\rho}^{\gamma}/m_{\rho}$ and
$\mathcal{A}_{\mu}^{2}= \varepsilon_{\mu\alpha\beta\gamma}\epsilon_{X}^{\alpha}
\epsilon_{\psi}^{\ast\beta}p^{\gamma}_{\psi}/m_{\psi}$,
if the $J/\psi$ and $\rho$ are in the $S$-wave channel.\footnote{%
  The $D$-wave channel contribution is greatly suppressed by the factor
  $(|\vec{p}_{\rho}|/m_{X})^2$ and may safely be neglected.}
We thus make the ansatz
\begin{equation}\label{amprho}
  \mathcal{A}_{\mu}(X(3872)\to J/\psi \rho)\propto\mathcal{A}_{\mu}^{1}+
  g\mathcal{A}_{\mu}^{2},
\end{equation}
with the relative-weight factor $g$ to be fitted to experimental data.
Adopting the masses $m_{X}=3.8717~\mathrm{GeV}$, $m_{\psi}=3.0969~\mathrm{GeV}$,
$m_{\rho}=0.7753~\mathrm{GeV}$, $m_{\pi^{\pm}}=0.1396~\mathrm{GeV}$, and the
total decay width $\Gamma_{\rho}=0.1491~\mathrm{GeV}$ from
Ref.~\cite{Tanabashi:2018oca}, using the functional form
$\mathrm{BW}_{\rho}(p_{\rho}^2)=(p_{\rho}^2-m_{\rho}^2+i\Gamma_{\rho}
\sqrt{p_{\rho}^2-4m_{\pi}^2})^{-1}$,
and integrating out the $\pi^{+}\pi^{-}$ phase space numerically, we find the
$X(3872)$ decay distribution, $W_X(\theta)$, to have the same form as in
Eq.~(\ref{jsipol}), with $\theta$ now being the polar angle of the $J/\psi$
flight direction in the $X(3872)$ rest frame, and the polarization parameter
therein to be
\begin{equation}\label{amrho}
\lambda^{X}_{\theta}=\frac{f(1-R)}{2-f+R},
\end{equation}
where $R=\sigma^{X}_{00}/\sigma^{X}_{11}$ and
$f=(-0.56+1.28g+3.12g^2)/(13.7+30.6g+18.2g^2)$.
Finally, we determine $g$ by fitting to the distributions of the
$X(3872)\to J/\psi\pi^{+}\pi^{-}$ partial decay width in the
$\pi^{+}\pi^{-}$ invariant mass $m_{\pi\pi}$, normalized to unity, as measured
by CMS \cite{Chatrchyan:2013cld} in the range
$0.5<m_{\pi\pi}<0.78~\mathrm{GeV}$
and by ATLAS \cite{Aaboud:2016vzw} in the range
$0.28<m_{\pi\pi}<0.79~\mathrm{GeV}$.
We thus obtain $g=-0.51\pm0.10$ with $\chi^2/\mathrm{d.o.f.}=35.3/22=1.60$.
The goodness of the fit can also be judged from Fig.~\ref{Fitpipi}, which also
contains the predictions evaluated with either $A^{1}_{\mu}$ or $A^{2}_{\mu}$
alone.
The latter results are somewhat worse, yielding $\chi^2/\mathrm{d.o.f.}$ values
of $45.9/23=2.00$ and $80.0/23=3.48$, respectively.
Other realistic functional forms of $\mathrm{BW}_{\rho}(p_{\rho}^2)$ yield
very similar results, albeit with slightly larger $\chi^2/\mathrm{d.o.f.}$
values.
Inserting our fit result for $g$ in Eq.~(\ref{amrho}) and setting in turn
$\sigma^{X}_{00}=0$ and $\sigma^{X}_{11}=0$, we obtain the allowed corridor
$-0.066\le\lambda^{X}_{\theta}\le0.141$, where the lower bound $f/(2-f)$, upper
bound $-f$, and 0 correspond to totally transversely, totally longitudinally,
and unpolarized $X(3872)$'s, respectively.
Our result for $W_X(\theta)$ is new.
We caution the reader that the functional form of $f$ depends on the
$\pi^{+}\pi^{-}$ phase space integrated over, so that $\lambda^{X}_\theta$ does
depend on the experimental acceptance cuts applied.
This must be taken into account in the extraction of polarization parameters
from experimental data of $X(3872)\to J/\psi\pi^{+}\pi^{-}$.

In our NLO NRQCD calculations, we use the on-shell mass $m_c=1.5$~GeV and the
two-loop formula for $\alpha_s^{(n_f)}$ with $n_f=4$ active quark flavors.
As for the proton PDFs, we adopt the CTEQ6M set \cite{Pumplin:2002vw}, which
comes with asymptotic scale parameter $\Lambda_{\mathrm{QCD}}^{(4)}=326$~MeV.
We choose the $\overline{\mathrm{MS}}$ renormalization, factorization, and
NRQCD scales to be $\mu_r=\mu_f=\xi m_T$ and $\mu_{\Lambda}=\eta m_c$,
respectively, where $m_T=\sqrt{p_T^2+4m_c^2}$ is the transverse mass, and
independently vary $\xi$ and $\eta$ by a factor of 2 up and down about their
default values $\xi=\eta=1$ to estimate the scale uncertainty.

\begin{figure}
\begin{center}
\begin{tabular}{cc}
\includegraphics[width=0.48\linewidth]{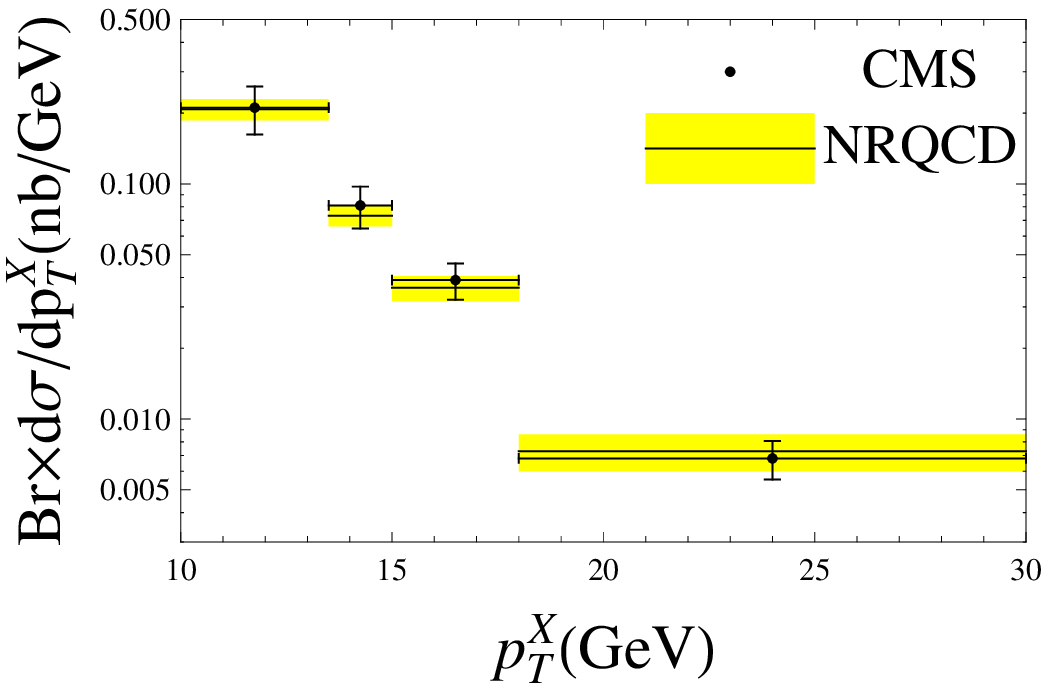}&
\includegraphics[width=0.48\linewidth]{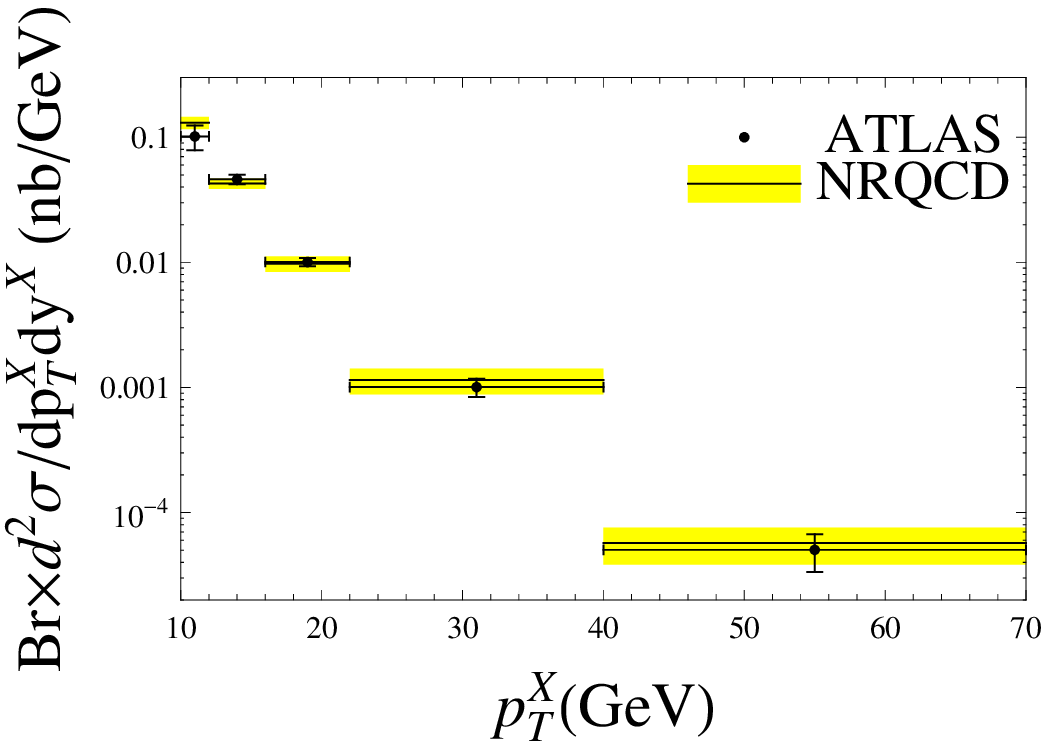}
\end{tabular}
\caption{The differential cross sections of prompt $X(3872)$ production as
  measured by CMS \cite{Chatrchyan:2013cld} (left panel) and ATLAS
  \cite{Aaboud:2016vzw} (right panel) are compared with our
  NLO NRQCD results based on the new fit (solid lines).
  The yellow bands indicate the theoretical uncertainties.}
\label{fitpt}
\end{center}
\end{figure}

The branching fraction $\mathcal{B}$ of $X(3872)\to J/\psi\pi^{+}\pi^{-}$ is
not yet known, so that we can only determine the products
$\langle\overline{\mathcal{O}}^{X}[n]\rangle\mathcal{B}$.
In our previous fit \cite{Butenschoen:2013pxa}, we included CDF
\cite{Acosta:2003zx,Bauer:2004bc}, LHCb \cite{Aaij:2011sn}, and CMS
\cite{Chatrchyan:2013cld} data of prompt $X(3872)$ hadroproduction.
Here, we perform an update by also including the recent ATLAS data
\cite{Aaboud:2016vzw}.
We obtain
$\langle\overline{\mathcal{O}}^{X}[{}^3P_1^{[1]}]\rangle\mathcal{B}
=0.34^{+0.12}_{-0.15}\times10^{-2}~\mathrm{GeV}^{5}$ and
$\langle\overline{\mathcal{O}}^{X}[{}^3S_1^{[8]}]\rangle\mathcal{B}
=0.83^{+0.12}_{-0.16}\times10^{-4}~\mathrm{GeV}^3$,
in good agreement with both our previous two-parameter fits
\cite{Butenschoen:2013pxa}, including and excluding the LHCb
\cite{Aaij:2011sn} data, lying in between them.
The fit quality is excellent, with $\chi^{2}/\mathrm{d.o.f.}=7.25/9=0.81$.
This is also evident from Fig.~\ref{fitpt}, where the cross sections of 
$X(3872)$ prompt hadroproduction, differential in $p_T^{X}$, as measured by
CMS \cite{Chatrchyan:2013cld} and ATLAS \cite{Aaboud:2016vzw}
are compared with our NLO NRQCD results.
Also the integrated cross sections
$\sigma^\mathrm{prompt}(p\overline{p}\to X(3872)+\mathrm{anything})
    \mathcal{B}=(3.1\pm0.7)~\mathrm{nb}$ and
$\sigma^\mathrm{prompt}(pp\to X(3872)+\mathrm{anything})
        \mathcal{B}=(4.26\pm1.23)~\mathrm{nb}$
measured by CDF \cite{Acosta:2003zx,Bauer:2004bc} and LHCb \cite{Aaij:2011sn}
are compatible with our respective NLO NRQCD results,
$(2.2\pm0.8)~\mathrm{nb}$ and $(5.8\pm1.5)~\mathrm{nb}$.
Here and in the following, the theoretical uncertainties are evaluated by
combining the scale and fit errors in quadrature.
Excluding the CDF \cite{Acosta:2003zx,Bauer:2004bc} and LHCb \cite{Aaij:2011sn}
data from our fit and so imposing the cut $p_T>10~\mathrm{GeV}$, we obtain
$\langle\overline{\mathcal{O}}^{X}[{}^3P_1^{[1]}]\rangle\mathcal{B}
=0.38^{+0.16}_{-0.20}\times10^{-2}~\mathrm{GeV}^{5}$ and
$\langle\overline{\mathcal{O}}^{X}[{}^3S_1^{[8]}]\rangle\mathcal{B}
=0.86^{+0.13}_{-0.19}\times10^{-4}~\mathrm{GeV}^3$,
with $\chi^{2}/\mathrm{d.o.f.}=3.83/7=0.55$.
Adopting these fit results instead would have an insignificant effect on the
predictions below.

\begin{figure*}
\begin{center}
\begin{tabular}{ccc}
\includegraphics[width=0.215\linewidth,angle=-90]{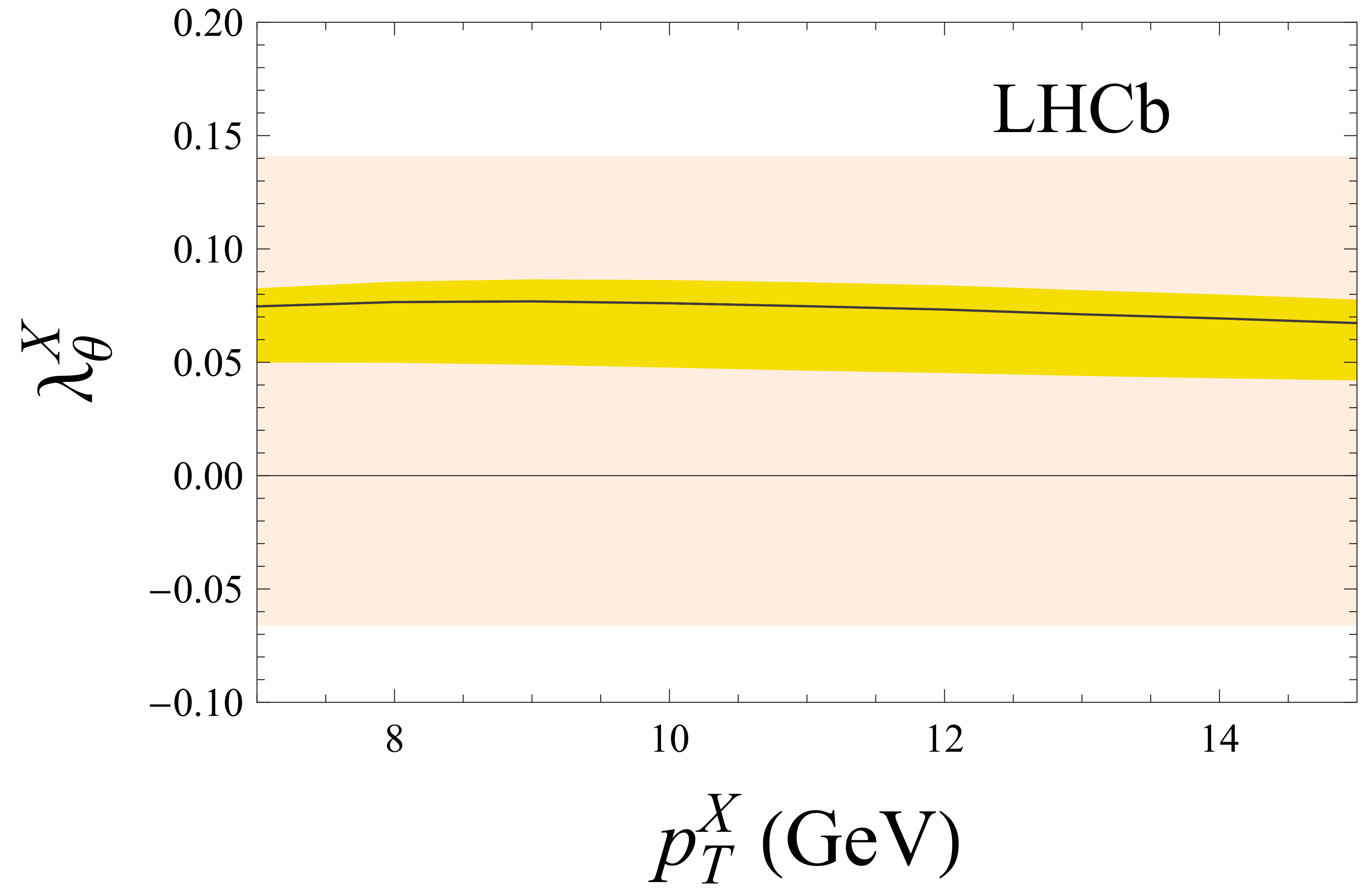}&
\includegraphics[width=0.215\linewidth,angle=-90]{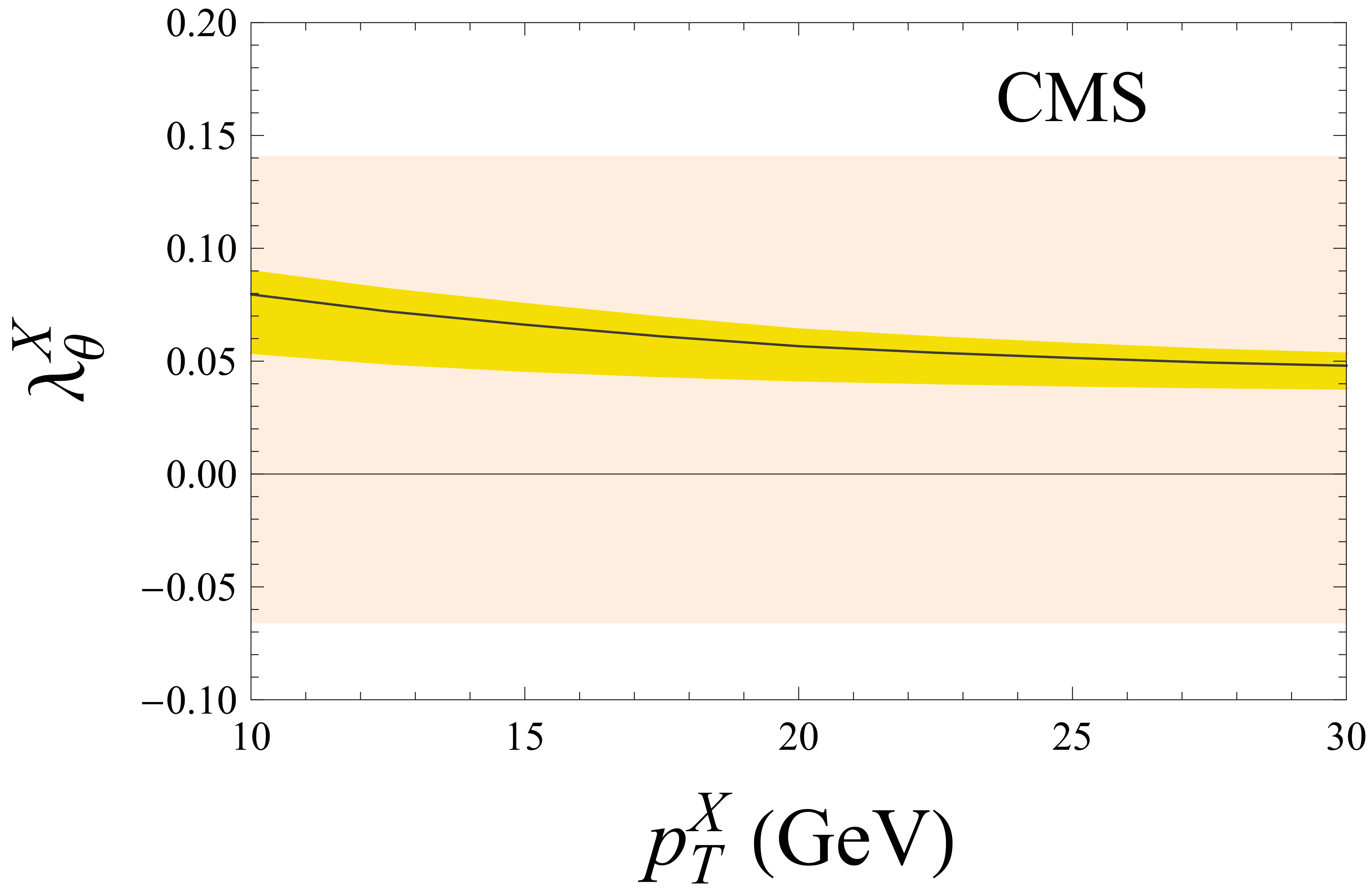}&
\includegraphics[width=0.215\linewidth,angle=-90]{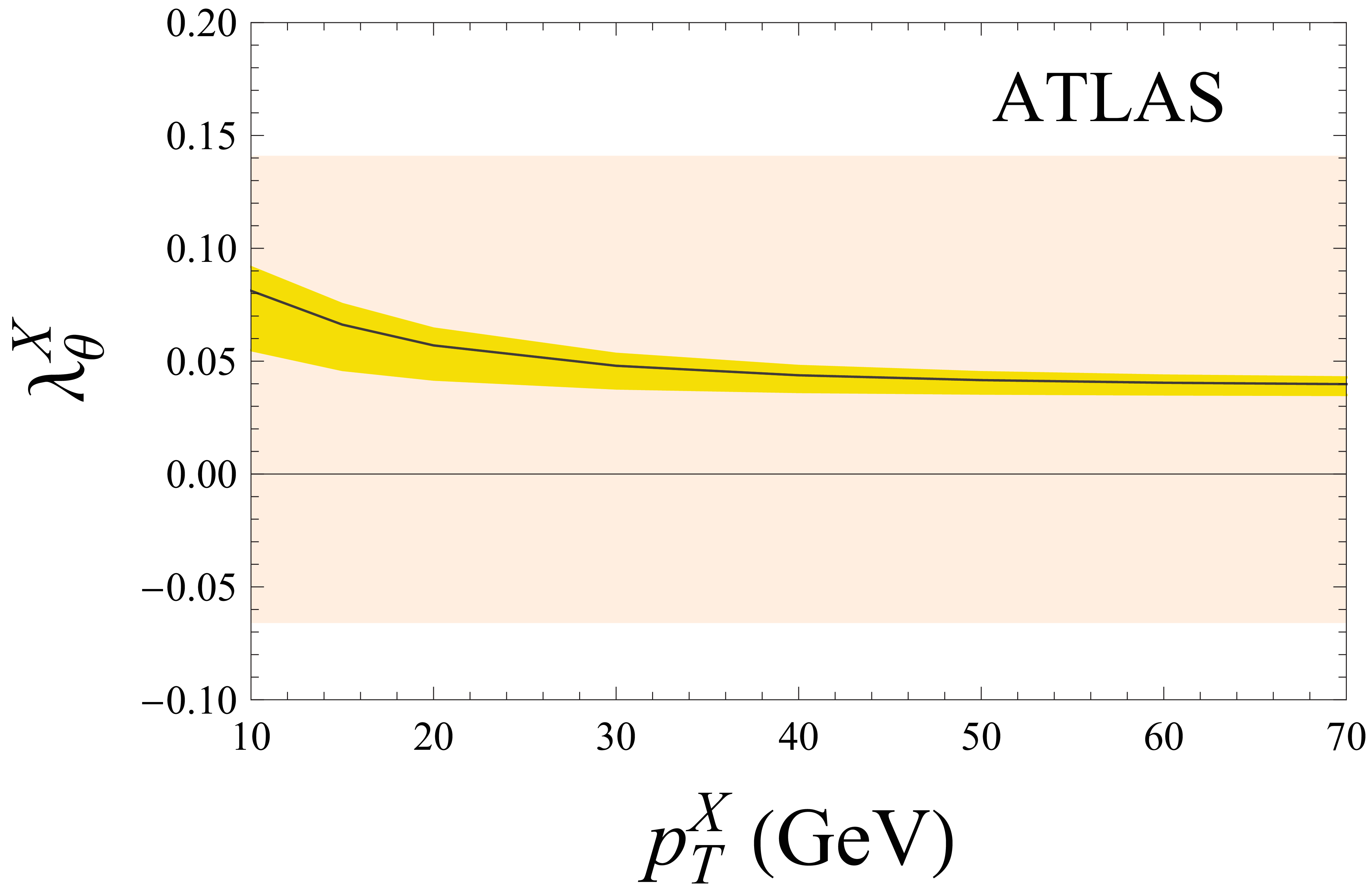}\\
\includegraphics[width=0.315\linewidth]{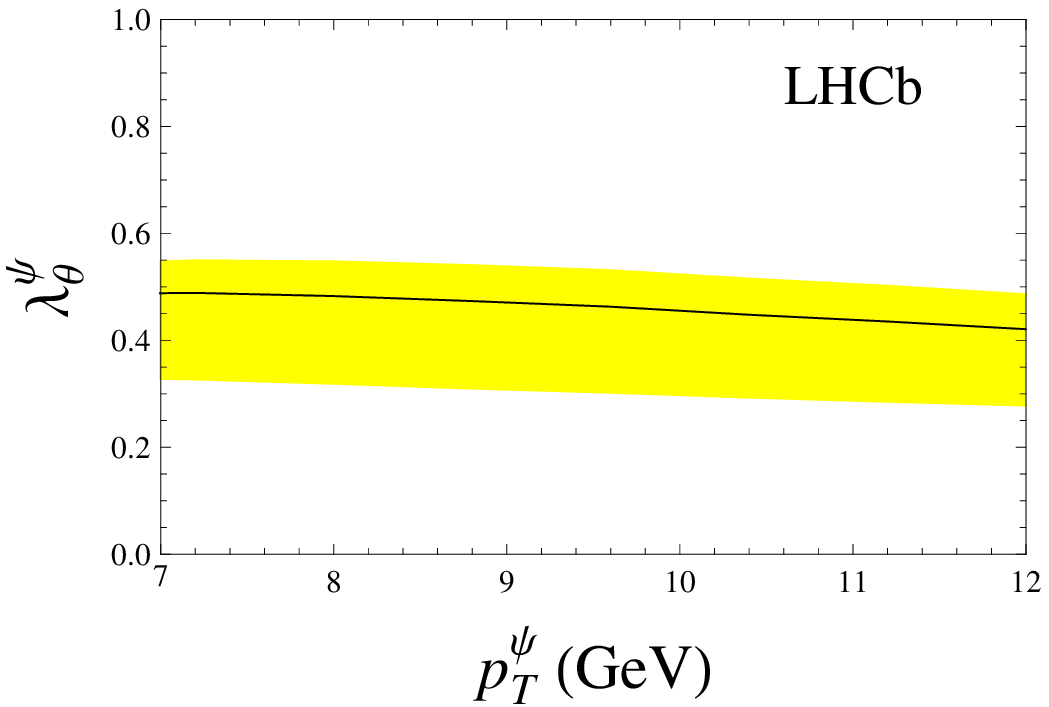}&
\includegraphics[width=0.315\linewidth]{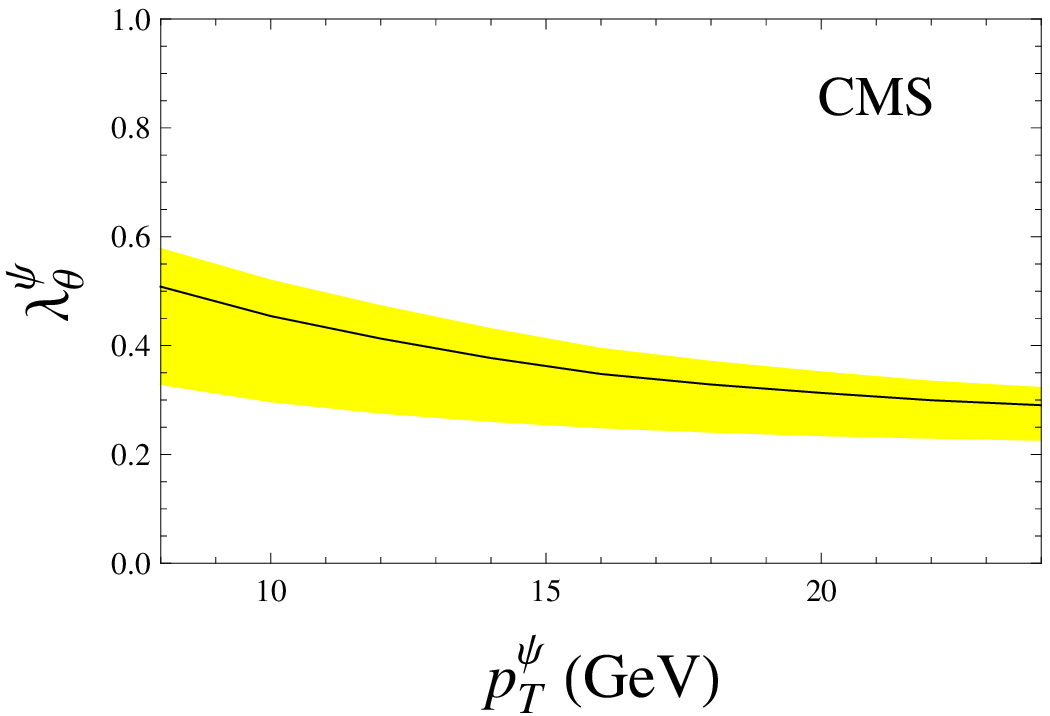}&
\includegraphics[width=0.315\linewidth]{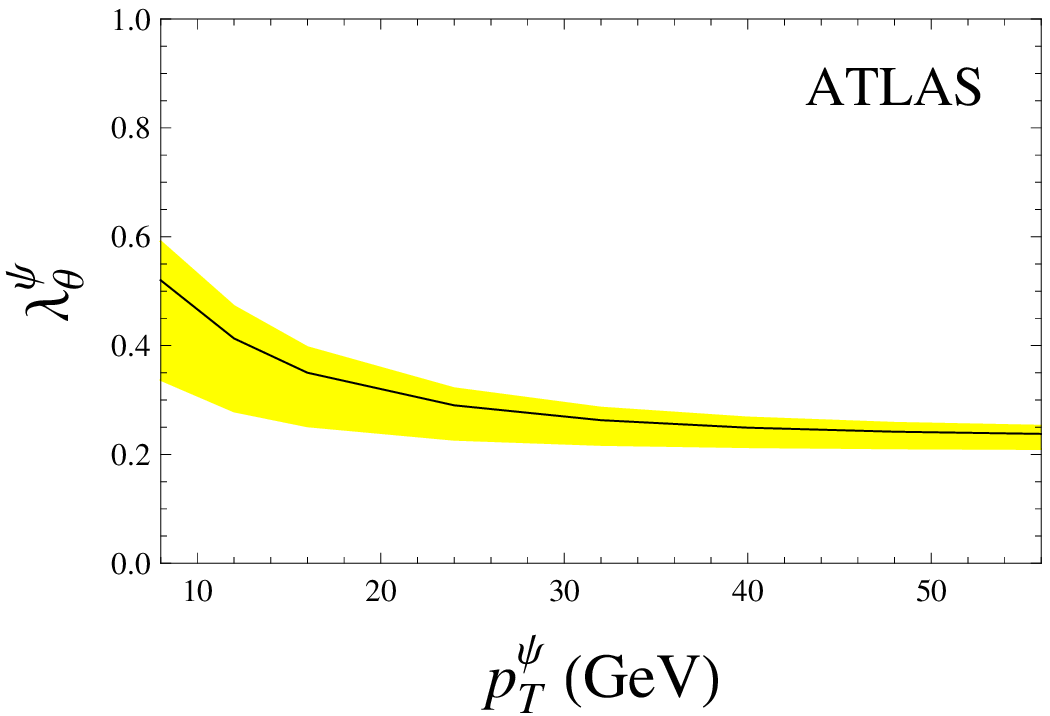}
\end{tabular}
\caption{$p_T^{X}$ dependence of $\lambda_{\theta}^{X}$ (upper row) and
  $p_T^{\psi}$ dependence of $\lambda_{\theta}^{\psi}$ (lower row) in the HX frame
  for prompt $X(3872)$'s decaying to $J/\psi\pi^{+}\pi^{-}$ for the LHCb
  (left), CMS (middle), and ATLAS (right) setups.
The shaded or pink bands indicate the allowed corridor for
$\lambda_{\theta}^{X}$.}
\label{Pol}
\end{center}
\end{figure*}

Imposing the $90\%$-C.L. lower bound $\mathcal{B}>3.2\%$
\cite{Tanabashi:2018oca}, we derive the upper bounds
$\langle\overline{\mathcal{O}}^{X}[{}^3P_1^{[1]}]\rangle
<0.11^{+0.038}_{-0.047}{}~\mathrm{GeV}^5$ and
$\langle\overline{\mathcal{O}}^{X}[{}^3S_1^{[8]}]\rangle
<2.6^{+0.38}_{-0.50}\times10^{-3}{}~\mathrm{GeV}^3$.
Since the factor $|\langle\chi_{c1}(2P)|X(3872)\rangle|^2$ cancels in the ratio
$r=m_c^2\langle\overline{\mathcal{O}}^{X}[{}^3S_1^{[8]}]\rangle/
\langle\overline{\mathcal{O}}^{X}[{}^3P_1^{[1]}]\rangle$,
we can also extract valuable information on the $\chi_{cJ}(2P)$ LDMEs,
\begin{equation}
  r=\frac{m_c^2\langle\mathcal{O}^{\chi_{c1}(2P)}[{}^3S_1^{[8]}]\rangle}
  {\langle\mathcal{O}^{\chi_{c1}(2P)}[{}^3P_1^{[1]}]\rangle}
  =\frac{m_c^2\langle\mathcal{O}^{\chi_{cJ}(2P)}[{}^3S_1^{[8]}]\rangle}
  {\langle\mathcal{O}^{\chi_{cJ}(2P)}[{}^3P_J^{[1]}]\rangle},
\end{equation}
where we have exploited heavy-quark spin symmetry relations valid to LO in
$v^2$ among the LDMEs for $J=0,1,2$.
This will in turn allow for new predictions of $\chi_{cJ}(2P)$ hadroproduction.
In this connection, it is interesting to observe that our central value
$r=0.055$ is consistent with the result for the $\chi_{cJ}(1P)$ case found in
Ref.~\cite{Ma:2010vd}, $0.045\pm0.010$.

With our new LDMEs, we are now in a position to predict $\lambda^{X}_\theta$
and $\lambda^{\psi}_\theta$ at NLO in NRQCD.
We consider three LHC setups as for $pp$ c.m.\ energy $\sqrt{S}$ and
$X(3872)$ rapidity $y^{X}$, corresponding to LHCb \cite{Aaij:2011sn}, CMS
\cite{Chatrchyan:2013cld}, and ATLAS \cite{Aaboud:2016vzw} experimental
conditions:
(i) $\sqrt{S}=7~\mathrm{TeV}$ and $2.0<y^{X}<4.5$;
(ii) $\sqrt{S}=7~\mathrm{TeV}$ and $|y^{X}|<1.2$; and
(iii) $\sqrt{S}=8~\mathrm{TeV}$ and $|y^{X}|<0.75$.
In Fig.~\ref{Pol}, $\lambda^{X}_\theta$ is presented as a function of $p_T^{X}$
for these three setups.
From there, we observe that the line shapes are very similar for the three
setups and that $\lambda^{X}_\theta$ ranges between 0.04 and 0.10, which is in
the upper part of the allowed corridor.
Thus, on the basis of our hypothesis, the $X(3872)$ is predicted to be
predominantly longitudinally polarized in all three setups, which is now up to
experimental verification.
Unfortunately, a reliable extraction of the $X(3872)$ polarization from the
measurement of $\lambda^{X}_\theta$ is hampered by the narrowness of its
allowed corridor, which necessitates high experimental precision.
We note that Eq.~(\ref{amrho}) is restricted to the decay channel
$X(3872)\to J/\psi\pi^{+}\pi^{-}$ and so is Fig.~\ref{Pol}.
However, the reader can easily extract the process independent quantity $R$
from Fig.~\ref{Pol} using the relationship
$R=(1+15.2\,\lambda^{X}_\theta)/(1-7.09\,\lambda^{X}_\theta)$ following from
Eq.~(\ref{amrho}) with our fit value of $g$.
$R$ is found to monotonically fall with increasing $p_T^{X}$, from
about 5 at $p_T^{X}=10~\mathrm{GeV}$ down to its asymptotic value of about 2.2.
Fortunately, the experimental disadvantage of $\lambda^{X}_\theta$ may be
circumvented by measuring $\lambda^{\psi}_{\theta}$ for the $J/\psi$ from
$X(3872)$ decay instead.
In fact, $\lambda^{\psi}_{\theta}$ is much more sensitive to the $J/\psi$
polarization than $\lambda^{X}_\theta$ is to the $X(3872)$ polarization.
Figure~\ref{Pol} also shows $\lambda^{\psi}_{\theta}$ as a function of
$p_T^{\psi}$ for the three setups.
From there, we observe that the $J/\psi$ is predicted to be largely
transversely polarized, especially in the lower $p_T^{\psi}$ range, where
$\lambda^{\psi}_{\theta}\agt0.5$.
Throughout the full $p_T^{\psi}$ range considered, $\lambda^{\psi}_{\theta}$ is
so distinctly separated from zero that this should be well discernible
experimentally with reasonable statistics.

In the molecular picture, $X(3872)$ is a loosely bound S-wave state of
$D^{\ast0}\bar{D}^0+c.c.$
Since $D^0$ ($\bar{D}^0$) is a pseudoscalar, all the information on the
$X(3872)$ polarization is carried by the $D^{\ast0}$ ($\bar{D}^{\ast0}$)
vector.
At hadron colliders, prompt $D^{\ast0}$'s ($\bar{D}^{\ast0}$'s) arise from the
nonperturbative evolution of perturbatively produced $c$'s ($\bar{c}$'s),
and we are not aware of a mechanism that leads to polarized
$D^{\ast0}$'s ($\bar{D}^{\ast0}$'s).
In fact, this argument is supported by several experimental measurements of
$D^{\ast0}$ ($\bar{D}^{\ast0}$) polarization in $e^{+}e^{-}$ annihilation at
different c.m.\ energies, for example by ARGUS
\cite{Albrecht:1996gr}.
We thus infer that, in the molecular picture, the prompt $X(3872)$'s
would be unpolarized and so would the $J/\psi$'s from their decays.

We now argue that the proposed $X(3872)$ and $J/\psi$ polarization measurements
are feasible using the LHC signal events already on tape by now and thus are
not subject to delay by the ongoing Long Shutdown 2, which will impede proton
physics before May 2021.
CMS managed to perfom a full-fledged $\psi(2S)$ polarization
measurement using 262\,k $\psi(2S)\to\mu^+\mu^-$ events in the range
$14<p_T^{\psi^\prime}<50$~GeV and $|y^{\psi^\prime}|<1.2$
\cite{Chatrchyan:2013cla}.
On the other hand, they collected 11.91\,k
$X(3872)\to J/\psi\pi^+\pi^-\to\mu^+\mu^-\pi^+\pi^-$ events in almost the same
kinematic range, $10<p_T^X<50$~GeV and $|y^X|<1.2$, using an integrated
luminosity of $4.8~\mathrm{fb}^{-1}$ \cite{Chatrchyan:2013cld}.
At present, the integrated luminosity accumulated by CMS is
$29.3~\mathrm{fb}^{-1}$ from run~1 and $160~\rm{fb}^{-1}$ from run~2
\cite{CERN}.
Assuming that acceptance and efficiency have been approximately steady during
the data taking periods, this translates into
$11.91\,\mathrm{k}\times29.3/4.8=72.7\,\mathrm{k}$ and
397\,k
prompt $X(3872)$ events waiting to be analyzed with regard to their $X(3872)$
and $J/\psi$ polarizations.
The data sample from run~2 alone is more than 50\% more copious than the one
underlying the $\psi^\prime$ polarization measurement \cite{Chatrchyan:2013cla}
and should thus conveniently allow for the proposed polarization measurements.
A similar conclusion can be drawn for LHCb on the basis of
Refs.~\cite{Aaij:2011sn,Aaij:2012ag}.

In summary, we studied the prompt hadroproduction of the mysterious $X(3872)$
and its subsequent decay to $J/\psi\pi^{+}\pi^{-}$ in the NRQCD factorization
framework \cite{Bodwin:1994jh} at NLO in
$\alpha_s$,
under the likely assumption that the creation of the $X(3872)$ proceeds
chiefly through the $\chi_{c1}(2P)$ component of its short-distance wave
function.
We updated our previous fits of the $X(3872)$ LDMEs \cite{Butenschoen:2013pxa}
by including the latest ATLAS data \cite{Aaboud:2016vzw}, scoring an excellent
goodness, as low as $\chi^{2}/\mathrm{d.o.f.}=0.81$ (see also
Fig.~\ref{fitpt}).
This also allowed us to predict the CO to CS LDME ratio of the
$\chi_{cJ}(2P)$'s.
Exploiting our fit results, we presented the first predictions of the
polarization parameters $\lambda^{X}_{\theta}$ and $\lambda^{\psi}_{\theta}$,
in the HX frame for LHCb-, CMS-, and ATLAS-like setups.
These imply that the $X(3872)$ and $J/\psi$ polarizations are largely
longitudinal and transverse, respectively.
Comparing the sizes of available LHC data sets on the $X(3872)$ prompt yield
with those on the $\psi^\prime$ polarization, we concluded that meaningful
measurements of $\lambda^{X}_{\theta}$ and $\lambda^{\psi}_{\theta}$ should be
feasible already now, during the LHC Long Shutdown 2.
While the reliable interpretation of such measurements of
$\lambda^{X}_{\theta}$ will be aggravated by the fact that the theoretically
allowed $\lambda^{X}_{\theta}$ window is only 0.21 wide, there is no such
limitation for $\lambda^{\psi}_{\theta}$.
Our predictions are distinctly different from those of the molecular picture
\cite{Close:2003sg}, in which the $X(3872)$ and $J/\psi$ vectors are
expected to be both unpolarized, as we argued on the basis of
Ref.~\cite{Albrecht:1996gr}.
Their experimental verification would simultaneously confirm both the validity
of NLO NRQCD for $\chi_{cJ}$ polarization and the $\chi_{c1}(2P)$ dominance
hypothesis in $X(3872)$ prompt production.
Should the $\chi_{c1}(2P)$ be discovered distinguishably from the $X(3872)$,
then the $\chi_{c1}(2P)$ dominance hypothesis could be tested, regardless of
the validity of NRQCD factorization, by comparing the $J/\psi$ polarizations
measured in $\chi_{c1}(2P)$ and $X(3872)$ decays.
We conclude by urging the LHC collaborations to extract $\lambda^{X}_{\theta}$
and $\lambda^{\psi}_{\theta}$ from available and future prompt-$X(3872)$ data and
to perform similar analyses also for other $X,Y,Z$ states with nonzero spin,
which will allow us to distinguish between different models and so to take a
major step in pinning down the nature of the $X,Y,Z$ states.

\begin{acknowledgments}
We thank Eric Braaten and Xiao-Rui Lyu for very useful discussions.
This work was supported in part by
BMBF
Grant No.\ 05H18GUCC1.
\end{acknowledgments}

\end{document}